# Current-induced electroresistance in $Nd_{0.5}Ca_{0.5}Mn_{0.95}Ni_{0.05}O_3$

A. Rebello and R. Mahendiran[*]

Department of Physics and NUS Nanoscience & Nanotechnology Initiative

(NUSNNI), Faculty of Science, National University of Singapore,

2 Science Drive 3, Singapore -117542

**Abstract**

We have investigated the *dc* and pulsed current-induced electroresistance in phase separated manganite $Nd_{0.5}Ca_{0.5}Mn_{0.95}Ni_{0.05}O_3$ (NCMONi05) as a function of temperature and magnetic field. It is shown that the negative differential resistance which appears above a threshold current ($I_c$) and hysteresis in the *V-I* progressively vanish with increasing period of the current pulses. However a strong non-linearity in *V-I* exists even for a pulse period of 6s. The peak voltage at $I_c$ decreases in magnitude and shifts towards higher current values with increasing strength of the magnetic field. The strong nonlinear behavior and the negative differential resistance in the *dc* current sweep are accompanied by a rapid increase of the sample surface temperature and therefore primarily arise from the Joule heating in the sample. While the Joule heating assists electroresistance in the high *dc* current regime, the nonlinearity in the pulsed current sweep and the resistivity switching between a high and low value induced by controlling the width and period of pulses can not be explained solely on the basis of Joule heating.

---

[*] Corresponding author R. Mahendiran email: phyrm@nus.edu.sg , Tel: +65-6516 2616, Fax: +65 6777 6126

Although perovskite manganese oxides (manganites) with the general formula $R_{1-x}A_xMnO_3$ where R = $La^{3+}$, $Pr^{3+}$ etc., and A = $Ca^{2+}$, $Sr^{2+}$ etc., shot into fame because of the discovery of colossal negative magnetoresistance in them in early 90's,[1] it is now clear that these materials are sensitive not only to the external magnetic field but to other external stimuli such as pressure, X-ray radiation, and electric field.[2] In particular, compounds which show coexistence of insulating charge-ordered (CO) and metallic ferromagnetic (FM) states show a dramatic variation in the resistivity by the application of such external stimuli. Among them, the current or electric field -induced collapse of the CO into the conducting FM state has been extensively investigated since its first discovery in $Pr_{0.7}Ca_{0.3}MnO_3$.[3] It was found that a large voltage ($V \sim 700$ V) is able to switch the insulating charge ordered state to a metallic ferromagnetic state due to the subtle competition between the CO and FM, and consequently, causes a huge decrease in resistance. Since the resistivity ratio $\rho(0)/\rho(V)$ can be several orders of magnitude ($10^2$-$10^8$ for the voltage amplitude 100-700 V), it was called colossal electroresistance (CER). Later, Guha et al.[4] found a strong non-linear voltage-current characteristic at a lower current ($I < 20$ mA). They have noted that the voltage increases linearly with current initially but above a threshold current ($I_{th}$), the voltage starts decreasing (i.e., negative differential resistance). These observations generated a lot of interest in the non linear electrical transport in these materials.

While many issues pertaining to the existence of CER in these oxides have been reported, its science has many puzzles and challenges. At present, there is no conscience about the unique origin of the electroresistance. In the beginning, the CER effect was observed below the charge ordering temperature in bulk $Pr_{0.7}Ca_{0.3}MnO_3$ and the origin of the CER effect was attributed to electric-field induced melting of charge order.[3] Later, the CER effect was observed above the charge ordered transition temperature or even at room temperature in a number of thin films of capacitance structure (metal-oxide-metal) where the oxide material is not necessarily charge ordered manganite but other oxides such as NiO, $SrZrO_3$, $La_{0.5}Sr_{0.5}CoO_3$, $La_{0.7}Sr_{0.3}FeO_3$ etc.[5] It might be possible that different mechanisms are operative in these systems which possess a wide range of electronic ground states. Many other mechanisms such as creation of conductive filamentary paths in the dielectric (charge ordered) matrix,[4] modification of Schottky barrier height at the interface,[6] trap controlled space-charge limited current,[7] excitation of charge density waves,[8] small-polaron-hopping,[9] phonon-assisted electron delocalization[10], Mott transition at the interface,[11] creation of defects and migration of oxygen ions [12] have also been proposed to explain CER but a coherent picture is yet to emerge.

There has always been some concern about the joule heating since significant joule heating at low temperature can lead to a decrease in resistance if the sample is semiconducting in nature ($d\rho/dT < 0$). The currents necessary for nonlinear effects are always on the limit of Joule heating and in this regard, distinguishing intrinsic mechanisms of the CER effect from those originating from thermal effects remains challenging and intriguing. More recently, Tokunaga et al.[13] studied the CER effect in phase separated $(La_{0.3}Pr_{0.7})_{0.7}Ca_{0.3}MnO_3$ and $Nd_{0.5}Ca_{0.5}Mn_{0.97}Cr_{0.03}O_3$ crystals. By using a magneto-optical imaging technique to visualize the inhomogeneous magnetization and conduction paths they demonstrated that the application of a large amount of current (~30-100 mA) can lead to the collapse of percolative conduction paths through a local Joule heating which leads to a large electroresistance. Current induced change in resistance accompanied by Joule heating was also reported by a few other researchers.[14] Although Joule heating can be significant if the sample is highly resistive and bad thermal conductor, it is premature to conclude that the CER effect is caused by the Joule heating alone. Extensive and unbiased investigations are necessary to elucidate the mechanism of the colossal electroresistance. In addition to the above mentioned *dc* current or voltage driven effects, there are reports on pulsed current or voltage-induced resistivity switching in manganites that are promising for application as non volatile memory (resistance random access memory).[15,16] This switching behavior occurs in such a way that the magnitude of electrical resistance between two electrodes changes after applying an electric pulse of different amplitude.

In this report, we emphasis on the *dc* and pulsed current induced electroresistance and magnetoresistance in a phase separated $Nd_{0.5}Ca_{0.5}Mn_{0.95}Ni_{0.05}O_3$ (NCMONi05). We present our results on the temperature dependence of the resistivity under various current amplitudes with the simultaneous measurement of the sample surface temperature. The switching of the resistivity in our sample with pulse period and pulse width as additional control parameters other than with the well known current amplitude is demonstrated. The effects of pulsed current and pulse parameters were not investigated in the related Cr-doped NCMO compounds reported previously.[7,8]

The temperature dependence of the resistivity and the *dc* and pulsed *I-V* characteristics were performed on a polycrystalline NCMONi05 sample of rectangular shape of dimensions 8 mm x 3 mm x 3 mm. The electrical contacts were made with Ag-In alloy and the distance between the voltage probes was fixed 6 mm. The sample was mounted on a standard thermal conductive puck and inserted in the sample space of a commercial superconducting magnet (Physical Property Measurement System). All the measurements reported here have been performed in four probe configuration using source-measure units (Keithley 2400 and Yokogawa GS 610 models). In order to investigate the change in temperature due to the Joule heating of the sample (*Tsurface*), a small calibrated Pt-100 resistance sensor (3mm x 2 mm x 1.2 mm in size) was thermally anchored to the top surface of the sample, between voltage contacts using the Apiezon-N grease and its four probe resistance during the current sweep was measured using a Keithley 6221 current source and a 2182A nanovoltmeter.

The main panel of Fig. 1 shows the temperature variation of the resistivity of the sample at three different current amplitudes ($I$ = 1 mA, 10 mA, and 20 mA). At a low current magnitude of 1 mA, the resistivity shows an insulator-metal transition around 60 K while cooling. The resistivity while warming is lower than while cooling in the temperature range from $T$ = 30 K- 95 K and then it crosses over the cooling curve over a certain temperature range before merging again at higher temperatures $T$ > 200 K. This insulator-metal transition is driven by the percolation of ferromagnetic clusters in the charge ordered insulating matrix similar to earlier observation in $Pr_{0.5}Ca_{0.5}Mn_{1-x}Cr_xO_3$ (x

= 0.015-0.05).[17] At higher current amplitudes $I \approx 10$ mA and 20 mA, the resistivity above 150 K matches with $I = 1$ mA data but they differ below 150 K. The insulator-metal transition and hysteresis seen in the $\rho$-$T$ curve at $I = 1$ mA disappear in the $\rho$-$T$ curves at $I = 10$ mA and 20 mA. Although $\rho$ for $I = 20$ mA is lower than the corresponding resistivity for $I = 1$ mA down to the lowest temperature from 150 K, the resistivity for $I = 10$ mA exceeds the resistivity at $I = 1$ mA below 50 K. Upon the application of a magnetic field $\mu_0 H = 5$ T, the insulator-metal transition is recovered and it occurs around $T = 100$ K for $I = 1$ mA and $T = 110$ K for $I = 20$ mA while cooling. The differences between the resistivity for $I = 1$ mA and 20 mA at 10 K under $\mu_0 H = 5$ T is almost negligible compared to the $\mu_0 H = 0$ T data. The inset shows the resistivity plotted against the surface temperature ($T_{surface}$) of the sample recorded by the Pt resistance thermometer thermally anchored to the sample surface. It is seen that the surface temperature of the sample does not go below 75 K when $I = 10$ mA (100 K for $I = 20$ mA) even though the base temperature recorded by the PPMS is 10 K. This suggests a strong evidence for Joule heating in the sample at higher current magnitudes.

The main panel of Fig. 2 shows the *dc* voltage-current (*V-I)* characteristics of the sample at selected stabilized PPMS base temperatures. Three consecutive current sweeps were performed, first sweep from 0 to +20 mA, second sweep from +20 mA to -20 mA and third sweep from -20 mA to +20 mA. The whole measurements were taken with a fixed current sweep rate of 0.5 mA/s and the initial surface temperature ($T_{surface}$) was ensured to be the same as the stable PPMS base temperature. For this we cooled the sample from 250 K to the respective base temperature before the start of each current sweep at different temperatures. While slower sweep rates than mentioned here has no impact on the *I-V* curves discussed in our results, faster sweep rates give rise to open and irreversible hysteresis loops. When the current is swept up from 0 to 20 mA at $T$ = 30 K, the voltage shows a sharp increase at lower currents followed by a peak around 10 mA before decreasing at higher currents. The decrease in voltage with current is well known in these types of manganites as the negative differential resistance (NDR) behavior.[4] In the second down sweep from 20 mA to -20 mA and third up sweep from -20 mA to 20 mA, the voltage-current curve traces different paths. Subsequent sweeps trace these curves except the initial one (0 to 20 mA). After a full cycle, a clear hysteresis is observed in the *V-I* characteristics. The *dc V-I* characteristics at other temperatures also exhibit NDR as well as hysteresis, but the hysteresis and strong NDR behavior disappears at higher temperatures as can be seen in the figure. The *V-I* characteristics is almost linear at 250 K and above. The top inset shows the resistivity as a function of current for a few selected temperatures. The resistivity change is negative and it often exhibits a butterfly-shape hysteresis with current. The inset at the bottom shows the change in the surface temperature (*Tsurface*) of the sample while sweeping the *dc* current. It is found that the observed non-linearity in *V-I* characteristics below 150 K is accompanied by a significant increase in the temperature of the sample. For instance, when the base temperature is 50 K, the surface temperature of the sample increases to as much as 110 K at the highest current $I$ = 20 mA. At higher base temperatures, the voltage-current characteristics become almost linear and a negligible change is observed in the corresponding surface temperature of the sample. The opening of hysteresis loop in *V-I* curves at lower temperatures needs a thorough study which is beyond the scope of this short communication. S. B. Roy et al. [18] investigated several charge-ordered manganites with quenched disorder and found that the high temperature charge ordered insulating phase is kinetically arrested during cooling and it can coexist with ferromagnetic metallic phase at low temperatures. Such a phase coexistence gives rise to unusual hysteresis loops (virgin curve being outside the envelope curve) in magnetization and magnetoresistance and can show a relaxor ferromagnetic behavior.[19] The magnitude of magnetization and magnetoresistance depends up on the different protocols used in the experiment, for instance field cool warm (FCW), zero field cool (ZFC), etc.[18,19] Since the temperature of the sample changes by several tens of kelvin during the current sweep in our samples, it is possible that the phase fraction of co-existing phases changes and in turn changes the resistivity.

Figure 3(a) compares the *dc* and pulsed *V-I* sweep at $T = 80$ K. The pulsed *V-I* sweeps were done with pulses of width, PW = 100 ms and five different pulse periods, PD = 200 ms, 500 ms, 1 s, 3 s and 6 s. It can be seen that the peak in voltage and NDR at higher currents observed in the *dc* current sweep disappear progressively with increasing pulse period. The magnitude of the voltage at the highest current keeps increasing with the pulse period (i.e., towards a high resistive state). The non-linearity in the *V-I* persists even for the longest period (PD = 6s). Figure 3(b) shows that the *dc V-I* sweep is accompanied by a significant increase in the surface temperature of the sample (~35 K), while the change in temperature decreases with increasing pulse period. The temperature change is only about 1 K for the pulsed sweep with PD = 6 s. The above difference in the voltage-current sweeps for the pulsed and direct current behavior suggests that caution should be taken while attributing the negative differential resistance to unpinning of charge density waves alone.[20] We believe that at a pulsed current of shorter pulse width or longer pulse period, intrinsic mechanisms possibly dominate over the thermal effects. Since the sample is phase segregated, the observed non linearity for the longer pulse period could also arise from the tunneling of trapped electrons between ferromagnetic clusters embedded in the charge ordered matrix.

Next we show how the *V-I* characteristics are affected by an external magnetic field. The main panel of Fig. 4 shows the *dc V-I* characteristics of the sample at $T = 80$ K for $\mu_0 H = 0$, 1, 2, 2.5, 3 and 5 T. To ensure the initial surface temperature (*Tsurface*) to be the same as the stable PPMS base temperature, we cooled the sample from 250 K to the respective base temperature before the start of each current sweep in different magnetic fields. When $\mu_0 H = 0$ T, the voltage shows a peak around $I = 3$ mA during the initial current sweep (0 to +20 mA). A large reversible hysteresis occurs upon subsequent current sweep. The voltage peak decreases in magnitude and shifts to a higher current value with increasing *H*. The hysteresis also decreases with increasing *H*. The peak is no more visible within the measured current range for $\mu_0 H \geq 2.5$ T. The top inset shows the resistivity and the bottom inset shows the corresponding change in the top surface temperature of the sample in different magnetic fields. The temperature of the sample increases as much as 35 K above the base temperature during the current sweep in zero magnetic field but the change becomes smaller ($\Delta T \sim 3$ K) when $\mu_0 H = 2$ T and almost negligible for $\mu_0 H > 3$ T. These results indicate a close competition between Joule heating and magnetic field to determine the threshold current in the NDR behavior. As the applied magnetic field decreases the resistance ($R$), the power dissipation in the material ($P = I^2R$) decreases and hence more current is required to induce NDR. So the voltage peak shifts towards higher value of current under higher magnetic fields.

All the above results indicate that Joule heating is significant in the *dc* current sweep, nevertheless it can be systematically controlled by manipulating the period of the current pulse and provide information about the contribution of other mechanisms in electroresistance. Hence we investigated the resistivity switching with different periods and pulses at $T = 80$ K. Figure 5(a) shows the response of the resistivity (left scale) in response to six pulse trains with a fixed pulse period (PD = 0.5 s) and pulse width (PW = 25 ms) but with two different current amplitudes, I = 2 and 10 mA. Consecutive pulse trains consists of 400 pulses of $I = 2$ mA and 400 pulses of $I = 10$ mA. Arrows in the figure indicate the start of a new pulse train. The resistance shows an abrupt increase by 50 % when the amplitude of $I$ is decreased from 10 mA to 2 mA and remains nearly unchanged until another pulse of a larger amplitude ($I = 10$ mA) sets it to a low resistance state. The resistance can be again set to the high value by the application of smaller current amplitude ($I = 2$ mA). The surface temperature of the sample, shown on the right scale, changes periodically from 87 K for $I = 10$ mA to 82 K for $I = 2$ mA.

In Fig. 5(b), we show a resistivity switching in response to six pulse trains with a fixed current of lower amplitude ($I = 2$ mA) and pulse period of 200 ms but for two different pulse widths PW = 25 ms and 100 ms. Consecutive pulse trains consists of 400 pulses of PW = 25 ms and 400 pulses of PW = 100 ms. The resistance decreases by 25 % upon changing the PW from 25 ms to 100 ms and the temperature also changes by 3 K. Similarly, a change of 16 % is obtained in a pulse period induced switching between PDs 50 ms and 100 ms pulses of fixed PW 25 ms and amplitude 2 mA, as shown in Fig. 5(c). In this case, the temperature changes by 1 K. The observed switching between two resistivity levels with varying PW may offer advantage in certain cases over the conventional method of resistance switching with changing magnitude of the current.

At present, the exact origin of the pulse width controlled resistance switching is not clearly understood. It is possible that longer is the pulse width, the continuous supply of energy ($W = I^2 R t_{pw}$ where $t_{pw}$ is the pulse width) causes the self heating of the sample and thereby leads to a decrease in the resistance. So, changing the pulse width periodically leads to heating and cooling of the sample repeatedly which leads to transition between low and high resistance states. Similarly, if the period of pulses in a pulse train is short, the sample does not have much time to dissipate the heat and hence the temperature of the sample is still high. Hence, shorter period leads to lower resistance state than the longer period of pulse train for a fixed pulse width. However, the observed changes in the temperature in fig 5 (b) and 5 (c) are only about 1-3 K which is not uncommon in conventional transistor or diode based devices. While the minority carrier tunneling through the depletion region is the primary cause of the non linear conduction in the diodes and transistor, Joule heating is a side effect.

There are other non-thermal mechanisms which can also account for the observed effect. In electronically and magnetically inhomogeneous systems such as the NCMONi05, electrons are delocalized within ferromagnetic domains but localized in the charge ordered domains. While the electrical conduction primarily occurs through the percolating ferromagnetic domains below the semiconductor-metal transition, electron transfer between disconnected ferromagnetic domains in the semiconducting state occurs through tunneling into and out of the charge ordered domains which acts like traps. In general, there will be distributions in the potential well (depth) of the traps and hence different life times. For a fixed pulse period, when pulse width is longer, charge carriers find enough time to tunnel through the well or get excited out of the well and contribute to electrical conduction leading to decrease in the resistance as observed in fig. 5(b). Hsu *et al*.[21] obtained qualitatively similar pulse width dependent results in metal-$Pr_{0.7}Ca_{0.3}MnO_3$-metal device (two probe configurations) though the authors did not measure the temperature change or the pulse parameter dependent *I-V* characteristics. It was suggested that the density of excessive nonequilibrium electrons near the cathode of the device caused the free valence electrons in the metal oxides to be localized. If the pulse width is larger than the relaxation time constant, the density of the nonequilibrium electron package injected from the cathode region is small. The high field intensity at the cathode region delocalizes the valence electrons. Recently, a charge trap model was proposed by Rozenberg and collaborators.[22] According to Rozenberg *et al*. there is a variation of the concentration of oxygen vacancies near the interface ("interfacial domains"). Under the action of alternative pulsing, oxygen vacancies moves back and forth between central region ("bulk") and interfacial domains thus effectively doping the central region. Thus pulsed current switching modifies the carrier concentration and it can explain the pulsed current or voltage induced resistance behavior. Finally, even though our results suggest that the Joule heating is non-negligible in the compound investigated, thermally controllable electrical switching may find other applications such as current limiter. [23]

In summary, we have shown that *dc* current-induced colossal negative electroresistance at a high current density in phase separated Ni-doped $N_{0.5}Ca_{0.5}MnO_3$ is related to the self-heating of the sample. However, non linearity in *V-I* characteristics does not vanish even with a pulsed current of a long (6 s) pulse-period which suggests that intrinsic mechanism of non-thermal origin is operative. It is shown that resistance can be switched not only with the amplitude of the current but also by changing the period and width of pulses. Though possible origins of the observed effects were suggested, a more detailed investigation in future will be helpful.

**Acknowledgements:** R. M thanks the Ministry of Education (Singapore) for supporting this work through the grant AcRF/Tier1- R144-000-167-112.

**Figure Captions**

**FIG. 1**: (Colour online) Temperature dependence of the resistivity of NCMONi05 at different current amplitudes ($I$ = 1 mA, 10 mA and 20 mA) under $\mu_0H$ = 0 T and 5 T. The inset shows the resistivity in zero field for all three current plotted versus surface temperature of the sample recorded by a Pt sensor attached to the top of the sample.

**FIG. 2**: (Colour online) Main panel: The *dc* voltage-current characteristics of NCMONi05 at different temperatures. The top inset shows the resistivity and bottom inset shows the surface temperature of the sample as measured by the Pt-resistance glued to the top of the sample during the current sweep.

**FIG. 3**: (Colour online) (a) *V-I* characteristics at $T$ = 80 K for different periods (PD) of the pulsed current. The pulse width (PW) is fixed to 200 ms. The *dc* data is also shown. (b) The change in the sample surface temperature during the current sweep.

**FIG. 4**: (Colour online) Main panel: Variation of the *dc* voltage-current characteristics under different magnetic fields at $T$ = 80 K. The top inset shows the resistivity and the bottom inset shows the variation of the temperature of the sample during the current sweep.

**FIG. 5**: (Colour online) Resistivity switching at $T$ = 80 K due to (a) change in the amplitude of the current (b) change in the pulse width (100 ms to 25 ms) for a fixed current $I$ = 2 mA and pulse period of PD = 200 ms (c) change in the pulse period (100 ms to 50 ms) for a fixed $I$ = 2 mA and pulse width of PW = 25ms.

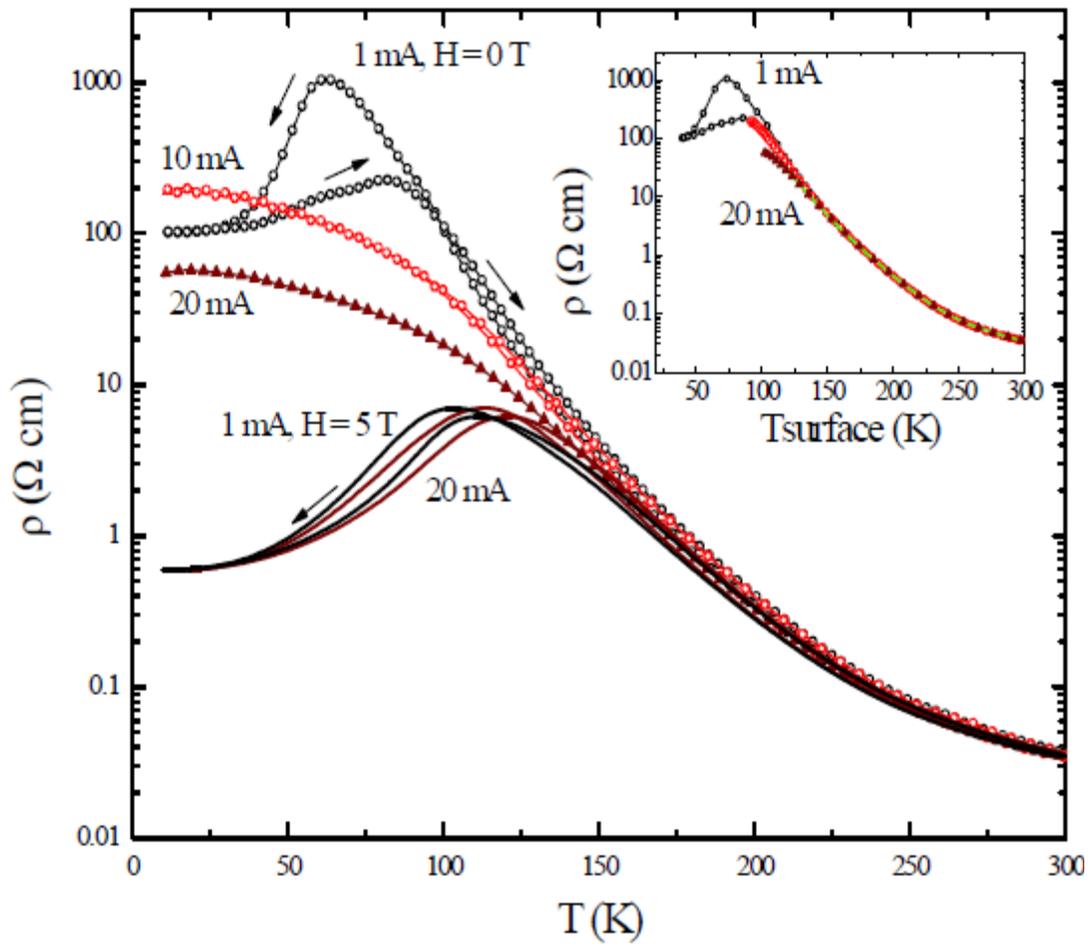

Figure 1

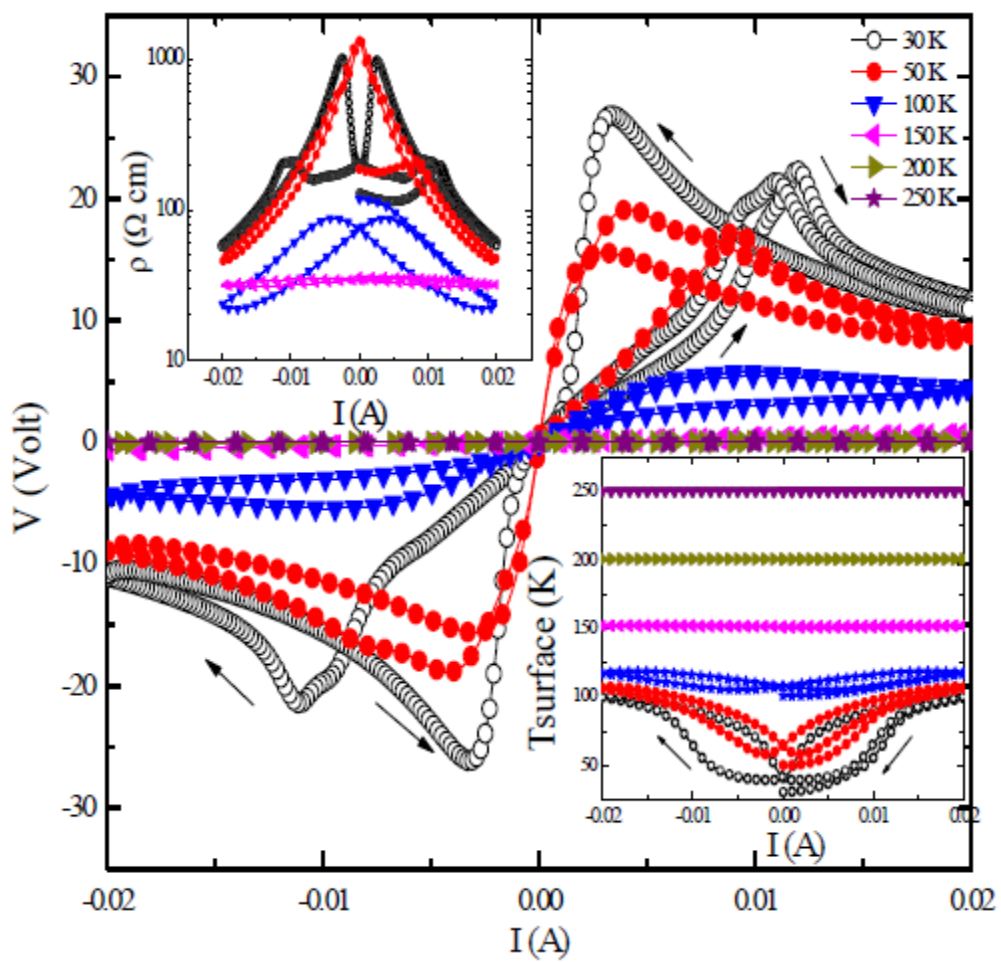

Figure 2

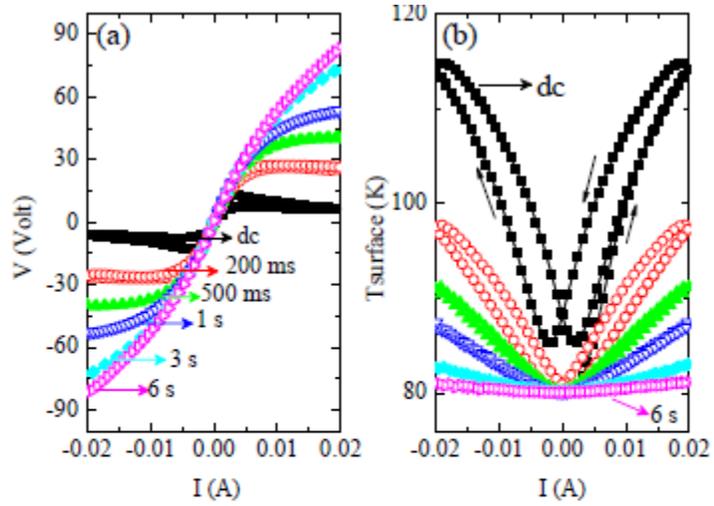

Figure 3

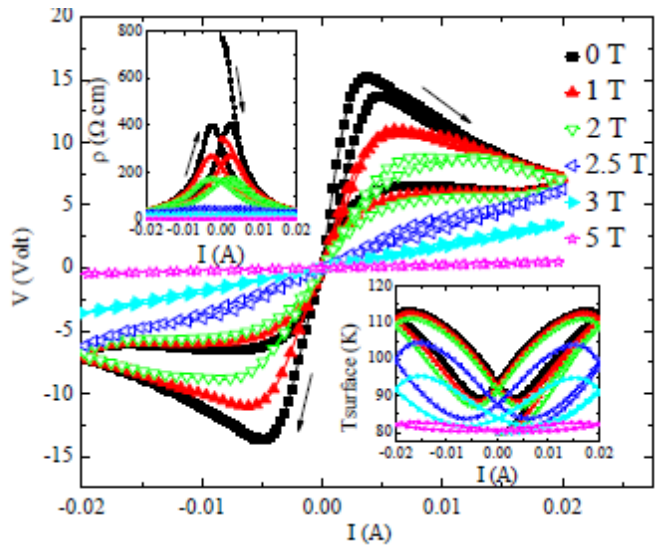

Figure 4

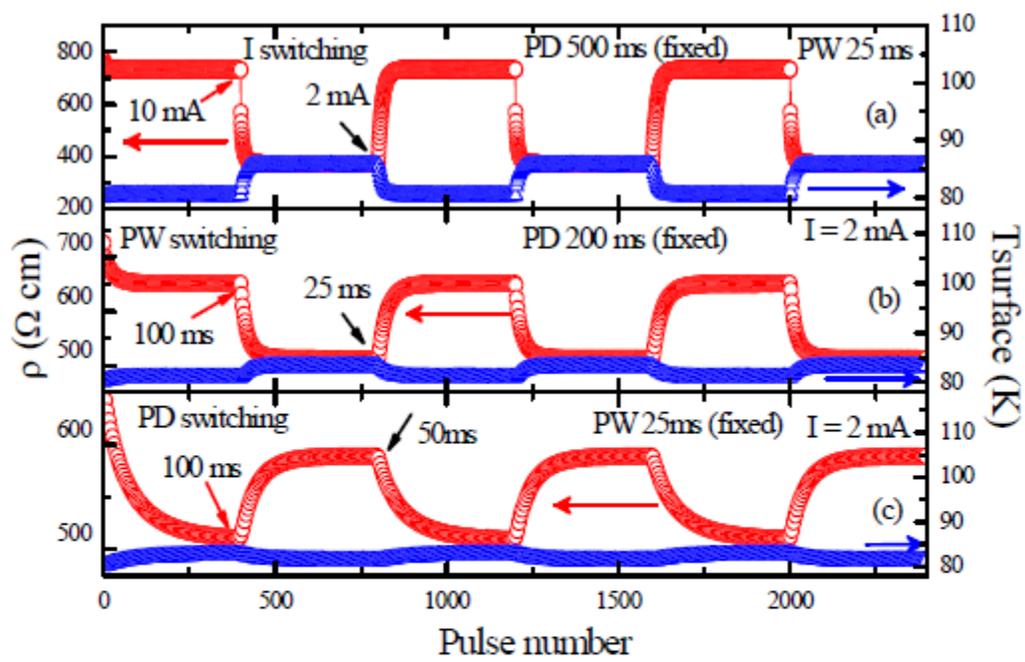

Figure 5